# Macro-level Indicators of the Relations between Research Funding and Research Output

Loet Leydesdorff [a] & Caroline Wagner [b]

**Abstract**

In response to the call for a science of science policy, we discuss the contribution of indicators at the macro-level of nations from a scientometric perspective. In addition to global trends such as the rise of China, one can relate percentages of world share of publications to government expenditure in academic research. The marginal costs of improving one's share are increasing over time. Countries differ considerably in terms of the efficiency of turning (financial) input into bibliometrically measurable output. Both funding schemes and disciplinary portfolios differ among countries. A price per paper can nevertheless be estimated. The percentages of GDP spent on academic research in different nations are significantly correlated to historical contingencies such as the percentage of researchers in the population. The institutional dynamics make strategic objectives such as the Lisbon objective of the EU—that is, spending 3% of GDP for R&D in 2010—unrealistic.

---

[a] Amsterdam School of Communications Research (ASCoR), University of Amsterdam, Kloveniersburgwal 48, 1012 CX Amsterdam, The Netherlands; loet@leydesdorff.net; http://www.leydesdorff.net .
[b] SRI International, 1100 Wilson Boulevard, Arlington, VA, 22209, USA, and George Washington University; cswagner@gwu.edu.

**Introduction**

In an editorial in *Science* on May 20, 2005, John H. Marburger III called for a "science of science policy." He formulated the challenge as follows: "How much should a nation spend on science? What kind of science? How much from private versus public sectors? Does demand for funding by potential science performers imply a shortage of funding or a surfeit of performers? […] A new 'science of science policy' is emerging, and it may offer more compelling guidance for policy decisions and for more credible advocacy." In response, an Interagency Task Group (ITG) was formed by the US National Science Foundation, the Department of Energy, and other agencies, in order to chart the domain. In November 2008, this ITG published a report containing "A Federal Research Roadmap" which elaborates Marburger's call in considerable detail and sets up a working program for the science of science policy (ITG, 2008).

In this study, we contribute to the discussion by providing recent data based on scientometric research at the macro-level of nations (De Moya-Anegón & Herrero-Solana, 1999; King, 2004; Shelton, 2008; Vinkler, 2005 and 2008). This data addresses some of the questions raised about the efficiency of spending in science. To this end, we bring together indicators that we developed in other contexts (Leydesdorff & Wagner, 2008, 2009; Zhou & Leydesdorff, 2006). However, nations may not be the appropriate units of analysis for scientometric analysis because one risks comparing apples with oranges. Average citation rates (and to a lesser extent average publication rates) differ among fields of science and even among specialties within fields of science. For example, papers in mathematics provide fewer citations than papers in the life sciences. Thus—as a policy implication—one might conclude that closing down a country's mathematics departments would improve its citation rate (Leydesdorff & Zhou, 2005).

Nations also differ in terms of funding schemes. For example, in some nations public R&D is concentrated in the universities and thus funded as part of the Higher Education Expenditure on R&D (HERD), while in other nations (e.g., China) the Academy represents a substantial part of the total R&D budget. Furthermore, macro-indicators by their very nature aggregate variety at lower levels of aggregation. At the level of nations, however, the prime concern is how the national research effort performs in comparison with other nations, in terms of contributing to national wealth, and in terms of converting financial resources in research output. Research output can be measured in terms of publications and patents. We limit ourselves in this paper to public spending and publications as indicators of output, but proceed by first positioning these indicators in relation to available indicators and other possible approaches. Our data suggests a number of conclusions, and raise questions for further research.

**Theoretical contexts**

The economic role of science & technology (S&T) entered the political agenda at the end of the 1950s when it became clear that economic growth could no longer be explained in terms of traditional economic factors such as land, labor, and capital. The remaining 'residue' has to be explained in terms of the emerging knowledge base of the economy (Abramowitz, 1956; OECD, 1964). Following the Sputnik shock of 1957, the *Organization for Economic Co-operation and Development (OECD)* was created in 1961 in order to organize and coordinate, among other things, science and technology policies among the member states, that is, the advanced industrial nations. In 1963, the first *Frascati Manual* was published in which parameters were defined for the statistical monitoring of science and technology on a comparative basis (OECD, 1963).

Output indicators of science emerged in the completely different context of information retrieval and the development of the *Science Citation Index*. During the 1950s and 1960s, the scientific community had itself become increasingly aware of the seemingly uncontrolled expansion of scientific information and literature during the postwar period (Garfield, 1955, 1978). In addition to its use in information retrieval, the *Science Citation Index* produced by the Institute of Scientific Information (ISI) soon came to be recognized as a means to objectify standards (Price, 1963; Elkana *et al.* 1978). The gradual introduction of output indicators (e.g., numbers of publications and citations) could be legitimated both at the level of society—because it enables policy makers and science administrators to use arguments of efficiency—and internally, because quality control across disciplinary frameworks is difficult to legitimate unless objectified standards can be made available in addition to the peer review process (Martin & Irvine, 1983; Moed *et al.*, 1985).

The econometric and the bibliometric approaches can both be made relevant for studying systems of innovation because innovations take place at interfaces between the knowledge production system and the economy (Rosenberg, 1976; Mowery & Rosenberg, 1979). In addition to industry-based systems of innovation (e.g., Nelson, 1982), the study of national systems of innovation was proposed by Freeman (1987), and elaborated by Lundvall (1988, 1992) and Nelson (1993). In this context, patent statistics are considered as indicators of inventions (Griliches, 1984, 1994; Narin & Olivastro, 1992; Pavitt, 1984; Sahal, 1981). Patent citations can be used for mapping the underlying science base of patents (Jaffe & Trajtenberg, 2002; Leydesdorff, 2004; Porter & Cunningham, 2005; Sampat, 2006; Sorenson & Fleming, 2004; Verspagen, 2006). Because of their legal and economic functions, patent citations can also be considered from perspectives different from the mapping of their knowledge base. For example, patent citations have been used for measuring the economic value of patents (Hall *et al*., 2002; Sapsalis *et al*., 2006; Trajtenberg, 1990). This has additionally been done at the level of innovation systems or even companies (Breschi *et al*., 2003; Engelsman & Van Raan, 1993, 1994; Leten *et al*., 2007).

Three literatures have emerged which are interrelated, but which have remained somewhat separate because of differences in their analytical perspectives: (*i*) the economic and econometric literature which focuses on the functions of R&D for growth in the economy, (*ii*) the bibliometric literature about indicators of growth in the sciences, and (*iii*) the literature in evolutionary economics which focuses, among other things, on patents as indicators at the interfaces between these two systems (Archibugi & Coco, 2005). In an authoritative study in the econometric tradition, Coe, Helpman, & Hoffmaister (2008, at p. 12) recently concluded that "there is robust evidence that total factor productivity, domestic R&D capital and foreign R&D capital are cointegrated [share a common trend, L&W], and that both measures of R&D capital are significant determinants of TFP [total factor productivity, L&W]." Four factors were found to make a difference among nations: (*i*) the ease of doing business, (*ii*) the quality of the tertiary education system, (*iii*) intellectual property protection, and (*iv*) the historical origins of the legal system. These conclusions, in our opinion, illustrate the extent to which the development of S&T itself is nowadays still black-boxed from this econometric perspective (Rosenberg, 1982): the qualification of labor, patent statistics, and R&D are considered as exogenous factors which can be expected to stimulate economic growth by interacting in an otherwise complex process.

The bibliometric approach and the perspective of evolutionary economics share a deeper appreciation of the dynamics of science and technology, and their respective literatures. In their groundbreaking book entitled *Patents, Citation, and Innovations: A Window on the Knowledge Economy*, Jaffe & Traitenberg (2002) considered patents and patent citations as providing data for studying the residue which cannot be explained in the econometric models of total factor productivity. The long-term impact of this residual knowledge factor—that is, R&D and education combined—on the economy is sometimes assessed to represent more than 30% of economic growth (Denison, 1964, at p. 54; cf. Mansfield, 1972, 1991, 1998; Romer, 1986; cf. Reikard, 2005). Analysts of patents and their citations focus on the decomposition of the databases in terms of industrial sectors (e.g., Grupp *et al*., 1996; Pavitt, 1984). With the shift to services induced by the knowledge-based economy, information technologies and open source innovations have become more important (Brécard *et al.*, 2006; Jorgenson & Stiroh, 1999; Von Hippel, 2001). S&T tend to operate globally, and therefore provide an additional dynamic of organized knowledge production and control to the economic system (Barras, 1991; Steinmueller, 2002). In a study of the German innovation system, for example, Leydesdorff & Fritsch (2006) showed that knowledge-intensive services and high-tech manufacturing are more "footloose" than medium-tech manufacturing in terms of infusing regions with knowledge.

The decomposition of organized knowledge production and control in terms of differences among the techno-sciences has been central to science and technology studies as a branch of the sociology of science (Whitley, 1984). Publication and citation practices, for example, are discipline-specific (Leydesdorff, 2008). However, under the pressure of the current institutional regime of university rankings, research evaluations, etc., institutional isomorphisms can also be expected which may cross disciplinary boundaries (DiMaggio & Powell, 1983). Given the availability of the *Arts & Humanities Citation Index*, for example, there are increasing pressures on researchers and research administrations even in the humanities to articulate instruments for comparative assessment (Steele *et al.*, 2006; Linmans, 2008).

Despite considerable efforts (e.g., Boyack *et al.*, 2005; Klavans & Boyack, 2009), the decomposition of the databases in terms of discrete clusters representing disciplines, specialties, interdisciplinary developments, etc., has hitherto remained an unresolved issue on the research agenda of bibliometric indicator studies (Rafols & Leydesdorff, 2009). What can nevertheless be said at the aggregate level in response to the questions raised by Marburger (2005) and the Roadmap (ITG, 2008)? How much should a nation spend on science? How does the emergence of China and other Asian nations impact on the science system and change the competitive relations among nations? (NSB, 2008)

**Methods and materials**

We use the latest available version of the *Main S&T Indicators* (2008-2) of the OECD (2008), and data from the *Science Citation Index* as available on the Web-of-Science of the Institute of Scientific Information (of Thomson Reuters). The latter are current data of publications and citations. However, we aggregate this data as yearly data using the tape-years of the ISI. (Publication dates may lag with reference to calendar years.) Consequently, 2008 is the last available year. The *Main S&T indicators 2008-2* are often incomplete for 2007, but almost completely available for 2006.

All expenditure data are normalized by the OECD in terms of current purchasing power parity, expressed in (millions of) US dollars. The best known among these indicators is GERD: Gross Expenditure on R&D. This indicator is often divided by GDP. As noted, the so-called Lisbon objective of the EU is to raise GERD to 3% of GDP in 2001. GERD is composed of three main components: Business Expenditure on R&D (BERD), Higher Education Expenditure on R&D (HERD), and Government Intramural Expenditure on R&D (GOVERD). Given the limitations of this study with its focus on publications, the latter two indicators are relevant.

HERD cannot be considered as a sufficient indicator of input to academic research because in some nations (e.g., China, the Russian Federation) the Academy is a major contributor to scientific publishing. In such cases, GOVERD is larger than HERD. With a focus on China, Zhou & Leydesdorff (2006) proposed to use (HERD + GOVERD) as relevant input for academic publishing, and BERD as an input indicator for patenting. However, the public research sector is often mission-oriented and therefore less driven by the institutional imperative to publish (Merton, 1942; OECD, 2002, at p. 41). We performed all analyses using both HERD and (HERD + GOVERD) as indicators for the input, and discuss the differences in the results. As noted, our focus is on the academic system and not on patenting, innovation, and business expenditure on R&D. Although one can argue for increasing integration among science, technology, and the economy (e.g., Narin *et al*., 1997), these integrations are not necessarily (and perhaps not primarily) taking place at the national level (OECD, 2007).

On the output side, we focus on the *Science Citation Index* because the capital intensity of the social sciences and the humanities is of a different order of magnitude. More specifically, it is best practice in scientometrics to include only citable issues, that is, articles, reviews, and letters. In October 2008, the ISI divided the document type "articles" into "articles" and "proceedings papers." This change was taken into account when collecting our data for 2008. Percentages of world shares of publications are based on whole number counts, but the value for the EU-27 was corrected for within-EU publications. The combination of the two databases enables us below to specify some of the dynamics at the level of nations.

The OECD statistics is organized assuming nations as units of analysis. In addition to statistics for OECD member states, information is provided for some non-OECD countries (e.g., China, Israel, the Russian Federation, and South Africa) which have adopted OECD definitions for gathering the statistics. However, one should remain aware that in a globalizing and increasingly knowledge-based economy, nations are not the only possibly relevant system of reference. Both markets and the sciences develop internationally. For policy analysis, however, nations have remained relevant units of analysis; nations function as systems by integrating different functions institutionally (Lundvall, 1988 and 1992; Nelson, 1993; Skolnikoff, 1993). The one exception to the latter statement, of course, may be provided by the EU. We use the EU-27 as an independent unit in the comparisons.

A consequence of the relative integration into a system at the national level is the expectation of possible spuriousness in the correlations among indicators. GDP may be regarded rather as a reason and not as a consequence of scientific development. Budgets are driven by institutional interests and by policy objectives derived from considerations such as international comparisons. The causality of a relation therefore is debatable and a multi-variate model is needed (Shelton, 2008). In other words, we don't claim any causality when specifying relationships between indicators in quantitative terms.

**Results**

Let us first focus on national outputs and then relate this to inputs.

*a. Output of nations in terms of publications*

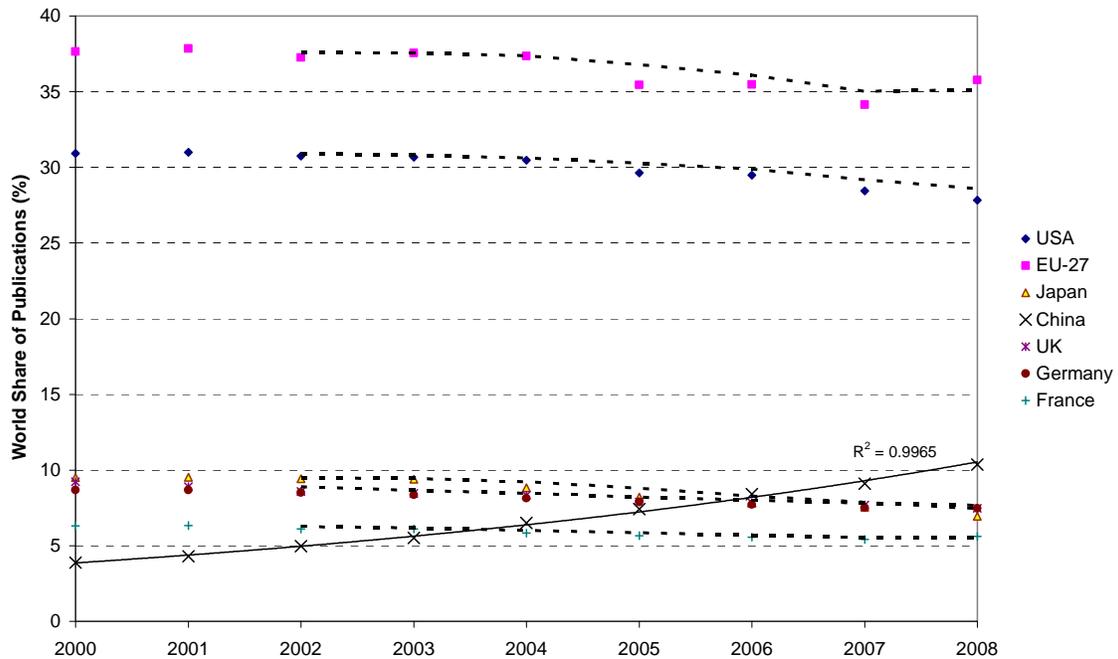

**Figure 1**: World Share of Publications for some major industrialized countries 2000-2008; based on articles, proceedings papers, reviews, and letters published in journals covered by the *Science Citation Index*.

Figure 1 shows the main output data of the science system in terms of publications, but normalized as a percentage of world share of publications for the period 2000-2008. In our opinion, the number of publications is the most unambiguous indicator of scientific output because no choices of parameters such as citation windows, etc., are involved. We added three-year moving averages as dashed trend lines, and the exponential regression curve for China ($r^2 > 0.99$). As previously, China continues to show exponential growth in both absolute and relative numbers (Zhou & Leydesdorff, 2006; Leydesdorff & Wagner, 2009). This leads to a relative decline for all other nations in the database because of the normalization involved in using percentages of world share.

As reported by Thomson-Reuters Scientific in May 2008, the database was recently extended in terms of "regional" coverage. This is visible in the above figures for the EU-27. The 2008-point reflects a rise from 3.0% to 4.4% for the twelve accession countries of the EU between 2007 and 2008. Some part of the increased share of China in 2008 may also be attributable to this one-time effect in coverage of the database. However, the main trends are clear: (*i*) the figures for China show a spectacular increase, and (*ii*) the USA is affected more negatively by the ongoing changes in the world system, but most industrialized countries have to give way (Shelton, 2008).

In a recent publication, we analyzed the citation patterns in more detail using, for example, the most highly-cited papers of the USA *versus* the EU (Leydesdorff & Wagner, 2009; National Science Board, 2008). In that study, we concluded that the USA tends to lose its dominance because the European countries are catching up, but this process of (mainly North-Atlantic) homogenization is relatively slow (Leydesdorff & Wagner, 2008). If current trends were to continue, China's contribution to world science could be as large as that of the USA by 2014.

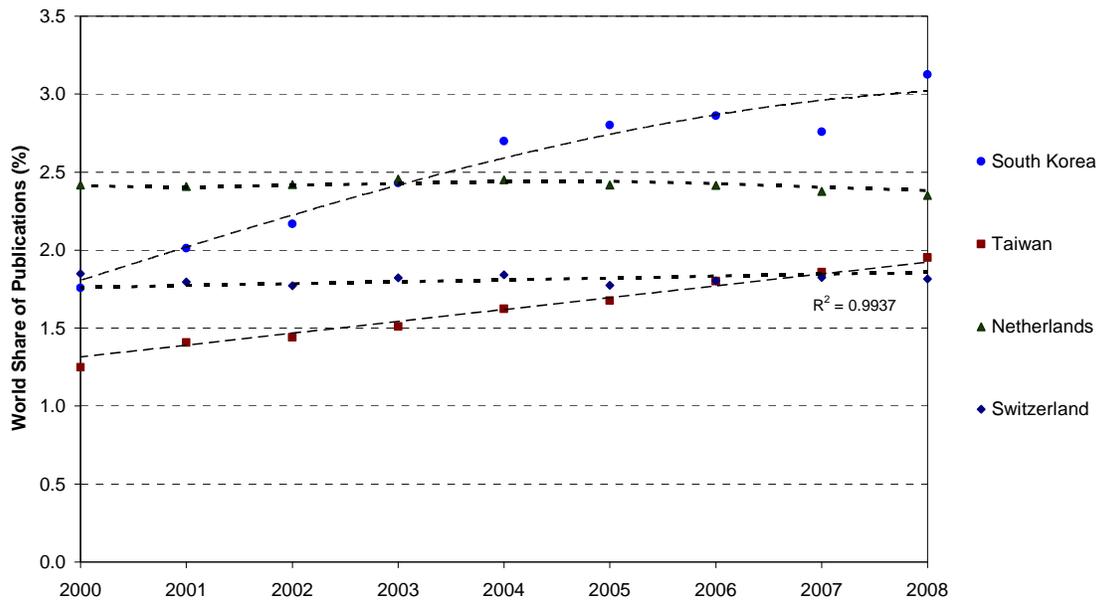

**Figure 2**: World Share of Publications for some smaller nations 2000-2008; based on articles, proceedings papers, reviews, and letters published in journals covered by the *Science Citation Index.*

Figure 2 shows the same trend lines for the Netherlands and Switzerland as two smaller, but R&D-intensive nations, and Taiwan and South Korea as two of the other so-called Asian-4 (including also Singapore with smaller numbers). The Netherlands is gradually losing in terms of world share of publications, but Switzerland seems to be able to keep its percentage of world share at the current level. Probably, the funding situation makes a difference between these two countries (cf. King, 2004).[1] South Korea shows a pattern of growth, but approximates its marginal percentage in the most recent years. The Scandinavian countries and the Netherlands also showed this pattern of linear growth with saturation in the 1980s, and Italy and Spain in the 1990s (Leydesdorff, 1992, 2000). This suggests that Korea is increasingly integrated into the world system. This is not yet the case for Taiwan, which is still able to convert domestic R&D into internationally oriented R&D output.

---

[1] The OECD statistics about Higher-Education Expenditure for R&D are unfortunately incomplete in the case of Switzerland (OECD, 2009: 40).

We suggest that the exponential curve in the case of China shows the effect of a positive feedback loop in the Chinese system. On the one hand, there is a large reservoir of academics eager to publish internationally and still to be tapped by the international system of publications; on the other, there are institutional incentives in place which stimulate international publishing in terms of financial rewards for individuals and research groups (Chen, 2005).

**b. Output / Input**

Nations are well-known to differ in terms of research portfolios, research intensity, and institutional structures. Japan, Sweden, and Finland, for example, spend more than 3% of GDP on R&D, while other nations with comparable levels of welfare (e.g., the UK, the Netherlands, and Norway) spend less than 2%. The composition of the national R&D budgets in terms of business, governmental, and higher education expenditure also varies among nations for both historical and economic reasons. Macro-indicators provide snapshots which cannot reflect nor explain detailed differences. However, they can serve us as a heuristic.

As noted, the OECD data is incomplete, but insofar as they are available all data are standardized in terms of current purchasing power parity (ppp), and the definitions of the indicators are well established. Furthermore, these definitions are used by national offices of statistics for the data gathering (OECD, 2002). Input statistics for S&T Indicators are provided by the OECD (2009) for 1995 and for the period 2003-2007.

Output measures are sensitive to the use of the database and many other decisions. Relevant citation windows, for example, vary among fields of science and publication types (Leydesdorff, 2008). Publication data using the ISI database and scientometric restrictions (such as only citable items), however, have increasingly been standardized in the scientific community. Since these two indicators are relatively straightforward, let us see what it brings us when we relate them.

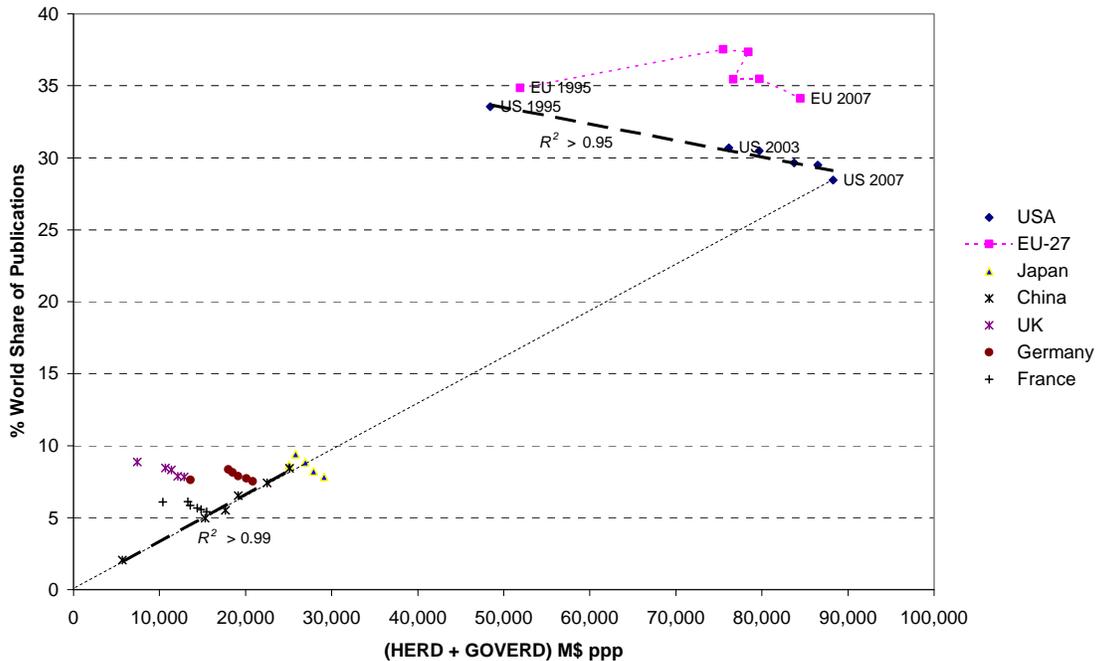

**Figure 3**: Output (% publications) over input (HERD + GOVERD) for major industrial nations and the EU-27 (1995, 2003-2007).

Figure 3 shows the relationship between the percentage of world share of publications *versus* the sum of HERD and GOVERD as costs. As is best visible for the US, countries tend to spend more over time to maintain their marginal percentage of world share of publications at a certain level. Note that these data are already corrected for inflation, and thus one could consider the additional spending as "R&D inflation" (cf. OECD, 2002, pp. 217ff.)  It seems that the EU-27 is less sensitive to this phenomenon than the USA (Shelton, 2008). The curve for the USA is equally declining when one normalizes on HERD only.

We added the linear regression line for the US in 2007 in order to enable the reader to see that the USA is spending per percentage point of output more than the European nations, but less than Japan. The data for China in the various years match this line ($r^2 > 0.99$). Figure 4 enlarges and details the lower left quadrant of Figure 3.

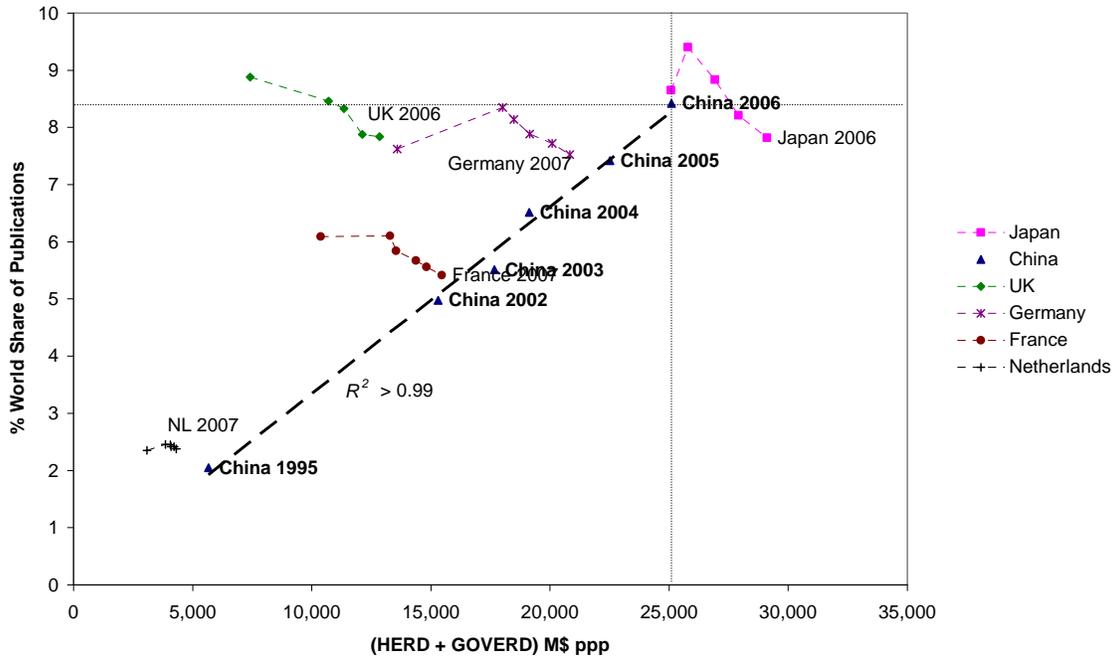

**Figure 4**: Output (% publications) over input (HERD + GOVERD) for several medium-sized nations and China.

All nations move with the years from the left to the right, that is, they increase their spending in R&D, and the trends are with the exception of China slightly downward, that is, they lose their percentage of world share as explained above. Japan economized on HERD during the 1990s and thereby (counter-intuitively) reached a higher percentage of world share of publications in 2003. However, Japan joined the club of comparable nations again in this millennium. Germany peaked in terms of output during the 1990s because of unification (Leydesdorff, 2000). France and the UK follow a similar pattern, but the UK is much more efficient in terms of output/input than any other nation. We added the Netherlands as an example of a small European nation.

The curve for China is penciled in as a regression line. In our opinion, this curve makes transparent what is happening in China: both expenditure and output are growing. China's efficiency (output/input) is now comparable with the OECD nations with the exception of the UK (and the Netherlands). The UK was traditionally highest in converting funding into publications. The Netherlands may be able to profit from spillovers of the Anglo-Saxon systems.

**c. How much should one spend?**

During their meeting in Lisbon in March 2000, the Heads of State of the European Union (the EU-15 at the time) and the European Commission agreed to raise the percentage of GDP to be spent on R&D to 3% in 2010. This objective can be considered as part of the so-called "Lisbon strategy" to make the EU the most competitive player in the knowledge-based economy (European Commission, 2000 and 2005). The underlying assumption that R&D investments will provide a return in terms of economic growth, however, has been debated, for example, in the case of the "Swedish paradox:" Sweden spent more than 3% of its GDP on R&D already during the 1990s. Furthermore, Sweden leads the OECD scoreboard in terms of various other indicators of knowledge-intensity (patents/ capita, new PhDs, etc.). However, its growth in GDP has remained below that of the OECD average to the extent that Sweden has moved from being one of the richest countries in the world, in term of GDP per capita, to a position below the OECD average (Edquist & McKelvey, 1998; Elg, 2003; Kander and Ejermo, 2008).

Note that from an econometric perspective, one can expect small countries like Sweden to profit from international R&D spillovers (Coe *et al*., 2008). The relatively low and still decreasing percentage (< 2%) that the Netherlands spends is often considered to be compensated by these effects of its open economy. Closer inspection of the OECD-data, however, reveals that relatively similar nations vary considerably in terms of the percentage of GDP they spend on R&D, and that these percentages tend to be rather stable over time. Historical factors such as the percentage of the workforce which is highly educated can be expected to have a long-term impact (Edgerton, 2008).

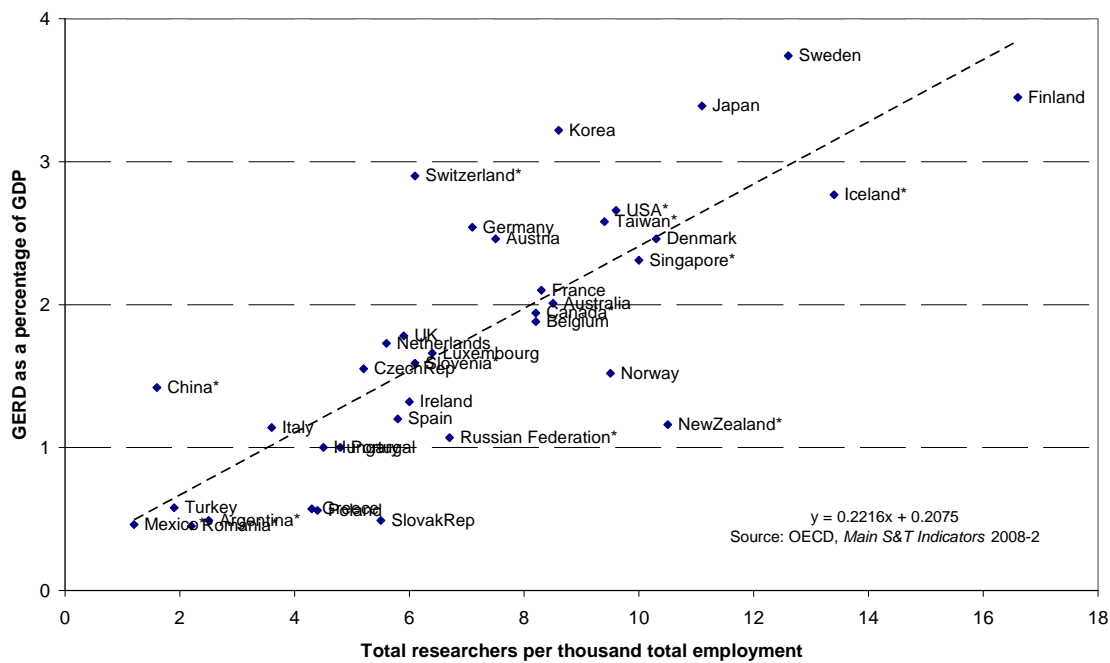

**Figure 5**: Percentage of GDP spent on R&D versus the permillage of researchers on total employment. Source: OECD, 2009.

In Figure 5, we plotted the %GDP spent on R&D against the total numbers of researchers per thousand total employment for a number of countries. The OECD (2009) provides 2006 data which can reliably be compared in the case of the 24 OECD member states. The two indicators correlate significantly: $r = 0.86$ ($p < 0.01$); $n = 24$. In the figure, we used the latest available figures for all countries listed (including a number of non-OECD countries), but asterisked the country names in the additional (non-2006) cases.

Horizontally, Figure 5 shows that countries are able to sustain very different workforces with similar funding efforts. New Zealand, for example, spends not more than Italy in terms of %GDP, but the percentage of its researchers in the workforce is more than three times as high. Along the diagonal, one witnesses the noted correlation. One can argue in the one direction that more government spending allows for more employment in this segment, or in the other direction that a larger share in the workforce corresponds to a stronger lobby for government spending in R&D. The mechanism is not known; the causality may also be circular.

In summary, these statistics suggest that the assumption under the Lisbon objective that government spending in R&D can be considered as a motor of economic growth is not warranted by the data. There is no direct short-term or long-term relation (as shown by the Swedish paradox), and similar countries can vary in terms of spending without producing a noticeable effect on their long-term economic performance. The harmonizing forces in the economy (for example, within the EU) can be analyzed in terms of international R&D spillovers, but this econometric reasoning cannot be turned around:

the harmonization in the economy as among the nation states of the EU has not led to similar patterns of R&D spending or to changes in the relative participation of the higher-educated segment of the workforce in the labor market.

Another perspective on the question of R&D funding can be provided by estimating a price per publication. Table 1 shows that the price per publication would be more than US$ 300,000 for the USA if normalized on HERD + GOVERD and approximately 175,000 when normalized on HERD only. For Ireland, for example, these figures are k$ 135 and k$ 107, respectively. For some Eastern European countries, they are even lower (Vinkler, 2008). When one compares France with Germany, one probably sees the difference in terms of governmental intramural R&D most sharply. German R&D is more concentrated in universities than it is in France.[2] Note that the cost per publication is higher for the EU-27 than for most of the EU-nations separately. This is due to the additional funding of R&D by the European Committee which cannot be associated with a separate component in the national outputs.

| 2007 | $ ppp/publication normalized on (HERD + GOVERD) | $ ppp/publication normalized on HERD | HERD as % of (HERD + GOVERD) |
|---|---|---|---|
| Australia* | 213,120 | 137,798 | 64.7 |
| Austria | 254,512 | 209,165 | 82.2 |
| Belgium | 159,189 | 115,328 | 72.4 |
| Canada | 241,861 | 191,004 | 79.0 |
| CzechRep | 207,846 | 97,939 | 47.1 |
| Denmark | 186,167 | 148,556 | 79.8 |
| Finland | 205,651 | 141,476 | 68.8 |
| France | 290,895 | 156,271 | 53.7 |
| Germany | 282,208 | 153,373 | 54.3 |
| Greece | 142,573 | 100,054 | 70.2 |
| Hungary | 177,085 | 87,047 | 49.2 |

---

[2] Luxembourg is left out of consideration because the normalized price per publication using HERD + GOVERD is US$ 487,793 in this case. This is obviously an outlier: the Luxembourg government spends only 18% of its R&D budget in higher education, and the public research sector is not active in publishing.

| | | | |
|---|---:|---:|---:|
| Ireland | 134,629 | 106,923 | 79.4 |
| Italy* | 216,745 | 132,441 | 61.1 |
| Japan* | 393,215 | 237,973 | 60.5 |
| Korea* | 285,342 | 132,019 | 46.3 |
| Netherlands | 185,943 | 124,708 | 67.1 |
| Norway | 286,855 | 193,157 | 67.3 |
| Poland* | 155,707 | 70,954 | 45.6 |
| Portugal | 179,399 | 137,362 | 76.6 |
| SlovakRep | 133,554 | 55,320 | 41.4 |
| Spain* | 205,436 | 128,119 | 62.4 |
| Sweden | 199,045 | 154,456 | 77.6 |
| Turkey* | 189,951 | 154,728 | 81.5 |
| UK* | 170,199 | 123,110 | 72.3 |
| USA | 316,287 | 175,284 | 55.4 |
| EU-27* | 269,479 | 163,216 | 60.6 |
| China* | 281,260 | 89,638 | 31.9 |
| Argentina | 324,109 | 137,901 | 42.5 |
| Romania | 279,271 | 115,938 | 41.5 |
| Russian Fed. | 351,509 | 62,763 | 17.9 |
| Singapore* | 256,439 | 179,017 | 69.8 |
| Slovenia | 145,269 | 58,081 | 40.0 |
| Taiwan* | 291,414 | 111,104 | 38.1 |
| Israel | 152,813 | 108,408 | 70.9 |
| | 228,382 ± 67,283 | 132,077 ± 42,696 | 58.5 ± 17.5 |

**Table 1**: Cost per publication for various nations. Sources: Web-of-Science and OECD data combined.

The rightmost column of Table 1 provides the percentage of governmental R&D expenditure spent in higher education. This is on average 58.5 (± 17.5) %. Indeed, two thirds of these nations spend between 41 and 76% of the budget on institutions of higher education. Some of the western countries (including Turkey) spend more; China and Russia considerably less.

As noted, part of GOVERD can be mission-oriented research and therefore not as strongly under the institutional imperatives of publishing as academic research (Merton, 1942). Therefore, it remains debatable whether one should use the second or third column of Table 1 for the normalization. However, the range itself is interesting. For example, this range is k$ 209-255 for Austria and k$ 141-206 for Finland. These differences cannot be explained in terms of the economies of the respective systems.

**Conclusions and discussion**

Nations differ in terms of their research portfolios and their economic structures. Both research fields and industrial structures are internationally organized. This raises a number of control problems at the national level. International scientific collaboration can be considered as leading to international R&D spillovers, but generates a new set of policy questions in the S&T arena (Wagner, 2008). From the perspective of hindsight, one can consider the OECD as an institution which, in reaction to the internationalization of the Western economies during the 1950s and 1960s, sought to monitor the effects of R&D on economic performance without opening the black box of science and technology itself. The transition from a (nation-based) political economy to a (globalizing) knowledge-based economy, however, makes opening the black box of organized knowledge production and control urgent.

The output statistics from the bibliometric perspective and patent statistics in evolutionary economics reveal other dynamics operating than the equilibrating ones of the market (Nelson & Winter, 1982). Research portfolios develop along historical trajectories and are entrenched in intellectual traditions that differ among nations and world regions. For example, while the USA is firmly committed to health research and has a strong pharmaceutical industry, China and other Asian nations are stronger in traditional fields like physics and chemistry (Park *et al*., 2005).

In our opinion, the discussion at the aggregate level of nations can no longer ignore the underlying complexities. The idea that innovation is coordinated in systems that are integrated at levels other than nationally, raises new questions for science and technology studies, indicator development, and econometric analysis. In a Triple Helix model, for example, governments are one of the partners in university-industry-government relations. Government policies can then be drawn into the analysis as both independent and dependent variables, depending on the research question and the stages in the development under study. National, regional, sectorial, and technological systems may compete in complex systems of innovations (Carlsson, 2006). The development of indicators for the extent of integration and synergy in different systems of reference is part of the research agenda (Leydesdorff, 2006).

Does the political discourse follow the developments in the economy and science and/or steer them (Dosi *et al*., 2006)? Questions about the dynamics of lock-in into and break-out from historical contingencies and traditions (Dolfsma & Leydesdorff, 2009) may prove more important in assessing returns on investments, than levels of spending at specific moments in time (e.g., the Lisbon objective of 3% of GDP in 2010). In our opinion, institutional incentives and constraints at lower levels of organization can be expected to prevail over national or transnational S&T objectives.

**Acknowledgement**

We are grateful to extensive comments by an anonymous referee on a previous version of this paper.

30